\documentclass[a4paper,11pt]{article}
\pdfoutput=1 % if your are submitting a pdflatex (i.e. if you have
\usepackage{jinstpub} % for details on the use of the package, please
\title{RPC application in muography and specific developments} %\boldmath A title with some math: $x=1$}

\author[a,1]{E. Le Menedeu,\note{Corresponding author.}}

\affiliation[a]{Universit\'e Clermont Auvergne, Universit\'e Blaise Pascal,CNRS/IN2P3, Laboratoire de Physique Corpusculaire, BP 10448, F-63000 CLERMONT-FERRAND,France}

\emailAdd{eve.lemenedeu@clermont.in2p3.fr}

\abstract{Muography is an imaging technique for large and dense structures as volcanoes or nuclear reactors using atmospheric muons. We applied this technique to the observation of the Puy de D\^ome, a volcano 2~km wide close to Clermont-Ferrand, France. The detection is performed with a 1m$\times$1m$\times$1.80m telescope made of 4 layers of single gap glass-RPCs operated in avalanche mode. The 1~cm$^2$ pad readout uses the Hardroc2 ASICs. The three data taking campaigns over the last three years showed that a RPC detector can be operated in-situ with good performances. Further developments to decrease the gas and power consumption and to improve the position and timing resolution of the detector are ongoing.}


\arxivnumber{1605.09218}%only if you have one

\collaboration[c]{on behalf of TOMUVOL collaboration}

\proceeding{XIII$^{\text{th}}$ Workshop on Resistive Plate Chambers and related detectors\\
  Feb 22 - 26, 2016\\
  Gent, Belgium}

\begin{document}
\maketitle
\flushbottom

\section{Context}
\label{sec:intro}

Volcanic hazard assessment and risk mitigation are two very important scientific subjects with heavy implications both on the population safety and economic development \cite{brown2015}. Anticipating future activity of volcanoes requires monitoring of their activity as well as information on their past-behaviour and their internal structure. Several geophysical methods are usually used to study the inner structure of volcanoes: electrical resistivity tomography\cite{revil2008}, gravimetry, that gives access to the density\cite{li1998, gailler2009}, magnetic survey, that images the local variations of the magnetic field induced by rocks\cite{gailler2012} or seismic tomography, that uses seismic velocity\cite{lees2007}.

These methods  are rather complex to interpret and sometimes, as in the case of electrical resistivity and gravimetry tomographies, the measurements need to be performed on the volcano itself, that is in a high risk area in case of active volcanoes. Moreover, they have low sensitivity to large depths and the inverse problem is often ill-posed.

Imaging with atmospheric muons\cite{alvarez1970, nagamine1995}, referred to as muography in the following, was recently made possible by the development of reasonably priced, large area, high efficiency and high precision muon trackers\cite{tanaka2007, marteau2012, fehr2012, tanaka2014}.

\begin{figure}[htbp]
  \centering
  \includegraphics[width=\textwidth]{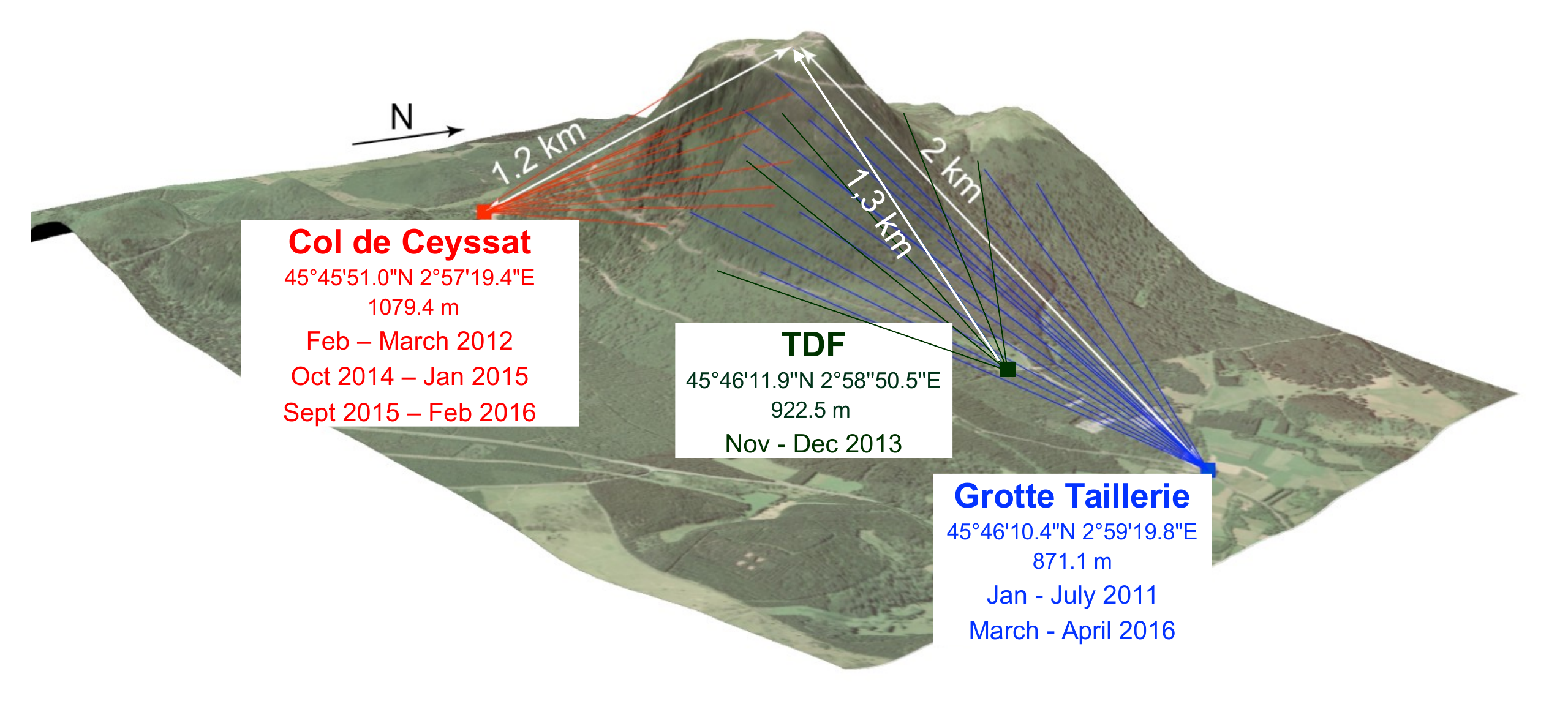}
  \caption{The Puy de D\^ome volcano and the locations where the detector has been deployed between 2011 and 2016.}
  \label{fig:PDD}
\end{figure}

Muography principle is the same as for radiography: the measurement of the absorption of a radiation through a target will give access to its transmittance image and its integrated density. Atmospheric muons are used here as they are naturally produced with a broad energy spectrum and can cross kilometers of rock before being stopped. This method is complementary to the other geophysical methods and offers clear advantages: it is a remote imaging, so the risk area can be avoided, it has a good spatial resolution and a well-defined inverse problem, at least in two dimensions. On the other hand muography suffers from the fact that high-energy muons are rare. This has a direct implication on the size of the detector to be used ($S_{det}$) and the duration of the campaigns ($\Delta T$). Indeed if a high resolution on density measurement is wanted, a high exposure ($S_{det}\times \Delta T \times\Delta\Omega$) will be needed (where $\Delta\Omega$ is the solid angle for viewing the target). %The exposure, in order to get 5~\% uncertainty on the density, is represented on the figure \ref{fig:expAelec} for the Puy de D\^{o}me volcano, supposing an average density of 1.8~g.cm$^{-3}$. 
In order to evaluate the feasability of muographic measurements, the TOMUVOL collaboration performed several data taking campaigns on the  Puy de D\^ome volcano, an averaged sized volcano, 2~km wide at the base, shown in figure~\ref{fig:PDD}. Assuming an average density of 1.8 g.cm$^{-3}$, as suggested by field and gravimetric measurements, the necessary exposure to measure the average density through muography with a 5~\% uncertainty is represented in figure~\ref{fig:expAelec} (left) as a function of elevation and azimuth. For example, a 5~\% uncertainty on density from an exposure of 1000~deg$^2.$m$^2.$day with a 1~m$^2$ detector can be achieved in one day for an angular resolution of 1000~deg$^2$, or in 1000 days for an angular resolution of 1~deg$^2$. The final result will be a compromise between the density resolution and the angular resolution, that will depend on the geophysical goal.

The TOMUVOL experiment takes place in this context: it is a proof of principle for imaging volcanoes with atmospheric muons using the Puy de D\^ome volcano, in the French Massif Central. It makes use of Glass Resistive Plate Chambers (GRPC) as they offer a large area for a reasonable price while providing highly segmented data. Three different campaigns have been recorded in two distinct locations: TDF 2013 and Col de Ceyssat in 2014-15 and 2015-16, as shown on figure~\ref{fig:PDD}.

\section{TOMUVOL detector}
The TOMUVOL detector is made of four layers of about 1~m$^2$, made of six GRPCs each. The chambers were built following the CALICE SDHCAL GRPCs\cite{beaulieu2015} design at IPNL\footnote{Institut de Physique Nucl\'eaire de Lyon, CNRS, Lyon 1}. Each GRPC is $50\times 33$~cm$^2$, in order to fit the geometry of CALICE's PCBs. The two glass electrodes are made of float glass 1.1~mm thick. The 1.2~mm gas gap is filled with 93~\% of TFE, 5.5~\% of isobutane and 1.5~\% of SF$_6$. The nominal high voltage is about 7.5~kV at normal pressure and temperature. It is corrected in real-time for $P/T$ variations.

\begin{figure}[htbp]
  \centering
  \includegraphics[width=0.45\textwidth]{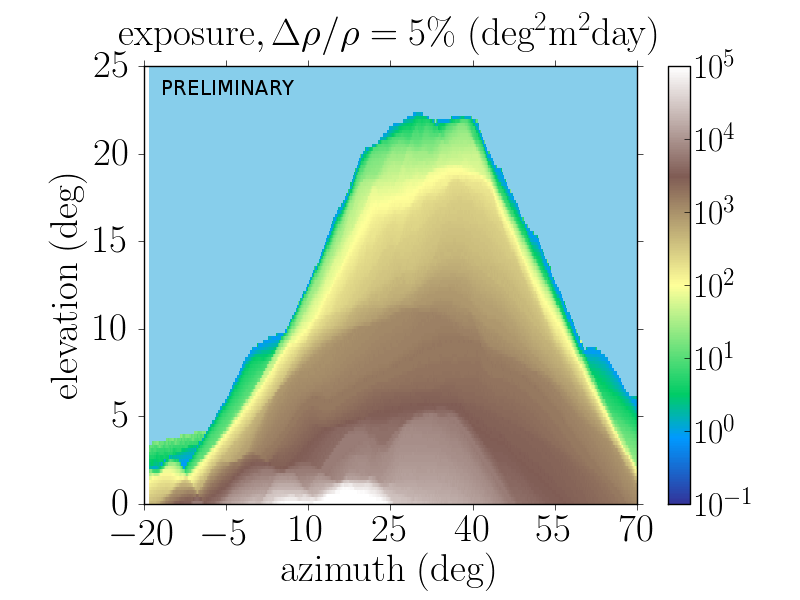}
  \hfill
  \includegraphics[trim=7cm 7cm 1cm 0cm, clip, width=0.45\textwidth]{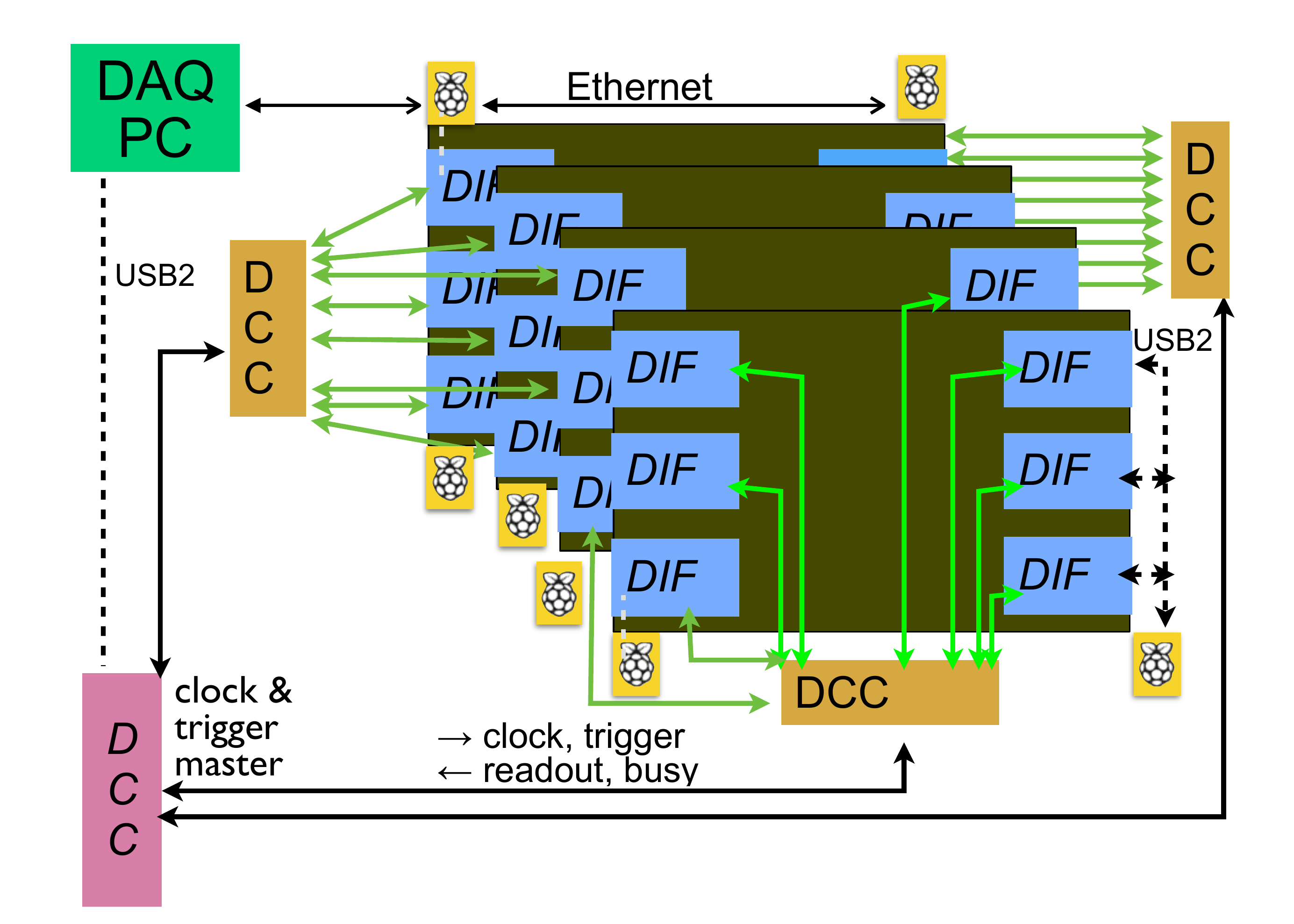}
  \caption{Left: exposure in order to get 5~\% uncertainty on the measured density. Right: schema of the TOMUVOL electronics.}
  \label{fig:expAelec}
\end{figure}

The GRPCs are read via pads of 1~cm$^2$, i.e. about 40000~pads in total, using low consumption ASICs (1.5~mW/channel when running continuously \cite{callier2014}) from Omega\footnote{Omega, CNRS, Palaiseau}, with 64~channels, semi-digital readout, following the SDHCAL design. The power budget is an important parameter for a telescope to be deployed on volcanoes : even in the rare cases when line power access exists, the available power is generally limited. The front end electronics is one DIF board from LAPP\footnote{Laboratoire d'Annecy-le-Vieux de Physique des Particules, Annecy} per chamber that also transmits the synchronous clock at 5~MHz, as represented on figure~\ref{fig:expAelec} (right). The detector is auto-triggered.

Slow control is performed via a PLC monitoring gas, low and high voltages and environmental conditions. It is remotely monitored from a web interface.

\section{Performances of the TOMUVOL detector}
\begin{figure}[htbp]
  \centering
  \includegraphics[width=0.565\textwidth]{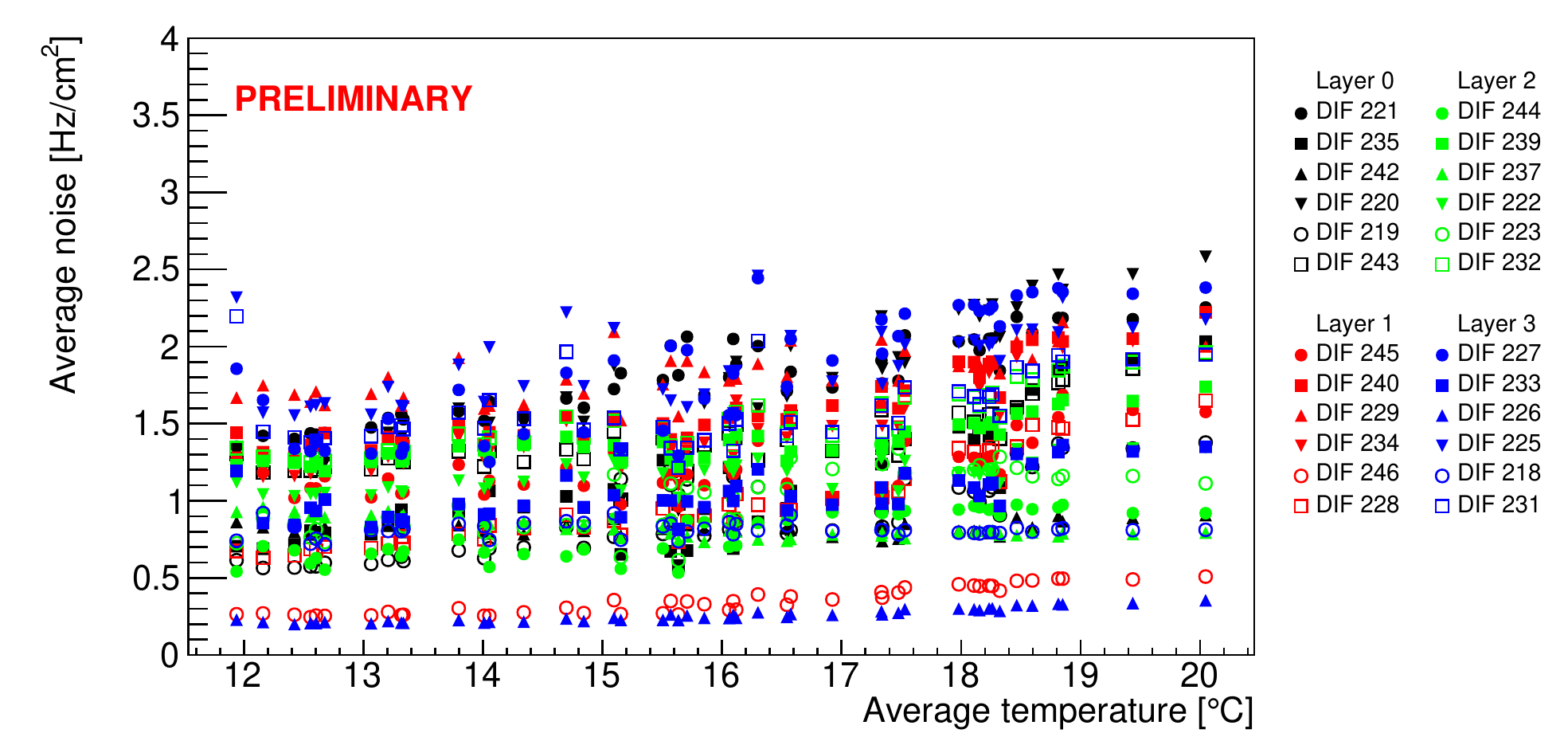}
  \includegraphics[width=0.39\textwidth]{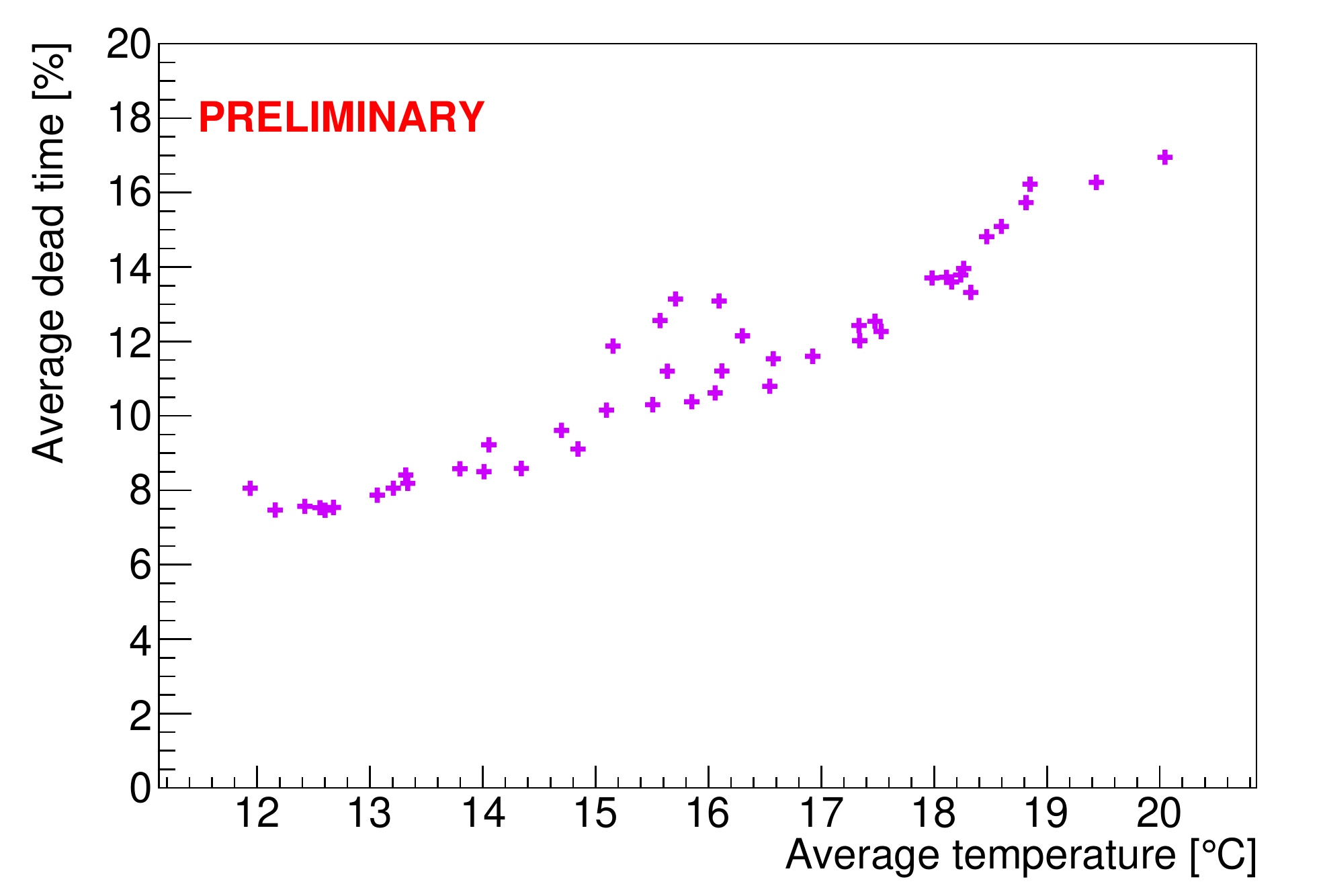}
  \caption{Mean noise (left) and dead time (right) in function of the temperature for the Col de Ceyssat campaign between December 10, 2015 and January 3, 2016.}
  \label{fig:correlTemp}
\end{figure}

%\subsection{Noise and dead time}
Being placed in-situ the TOMUVOL detector performances do vary with the atmospheric conditions. A dedicated study of the gain as a function of the environmental conditions was not yet performed. As a first approach, we followed the recipe suggested in \cite{gonzalez2005}, so high voltage is corrected for pressure over temperature variations through
\begin{equation*}
  HV_{eff} = HV \times f_{corr} \qquad \mathrm{with} \qquad f_{corr} = \frac{P}{P_{ref}}\frac{T_{ref}}{T} \qquad(P_{ref} = 1.013\ \mathrm{hPa}, T_{ref} = 293.15\ \mathrm{K})
\end{equation*}
but we observe some remaining correlation with temperature. They can be seen on figure~\ref{fig:correlTemp}, where noise and dead time are represented versus temperature. It has to be stressed that in our case the data aquisition is limited by the USB protocol used for reading the front-end boards, that means that dead time is increasing when noise is increasing.\\

\begin{figure}[htbp]
  \centering
  \includegraphics[width=0.4\textwidth]{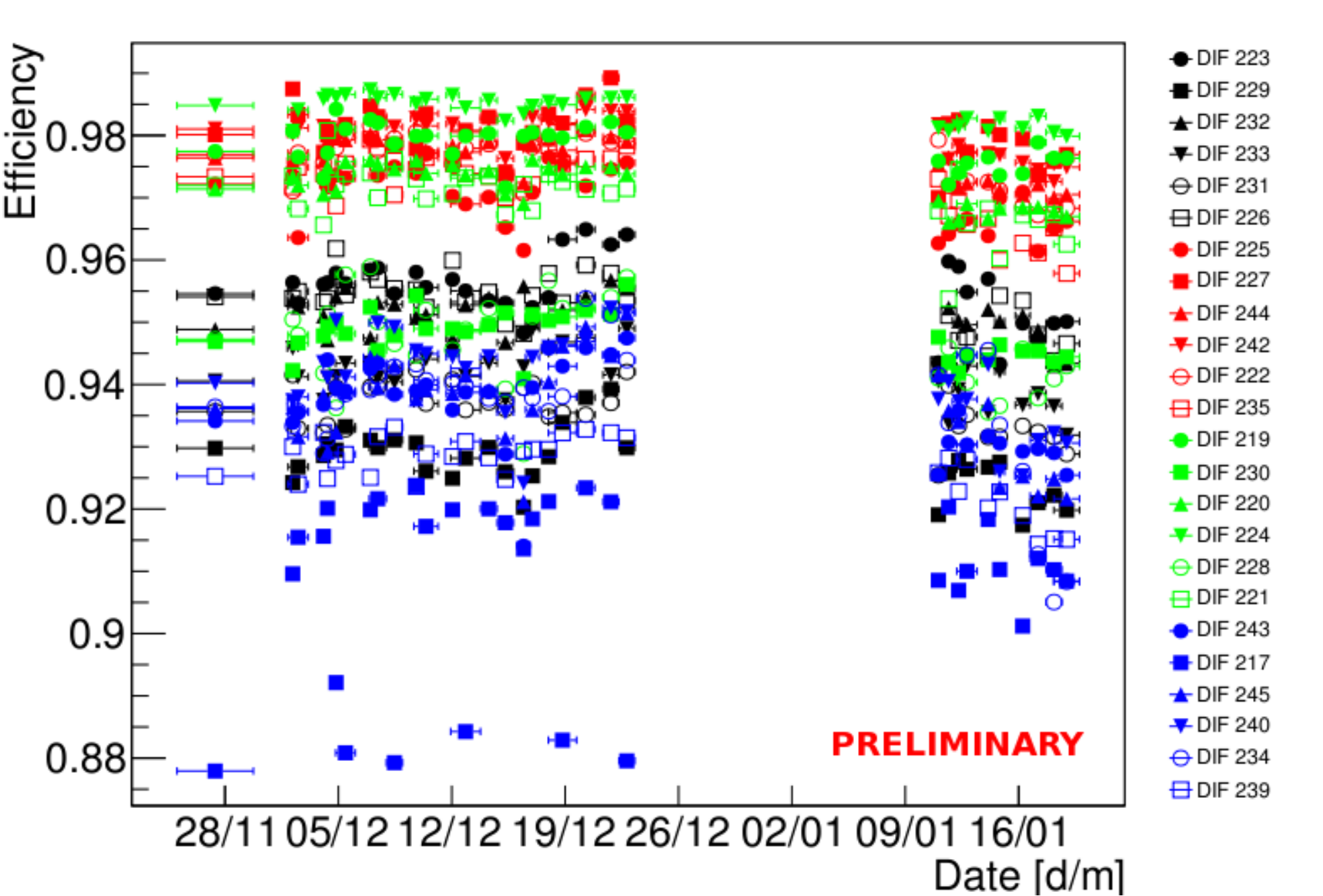}
  \includegraphics[width=0.5\textwidth]{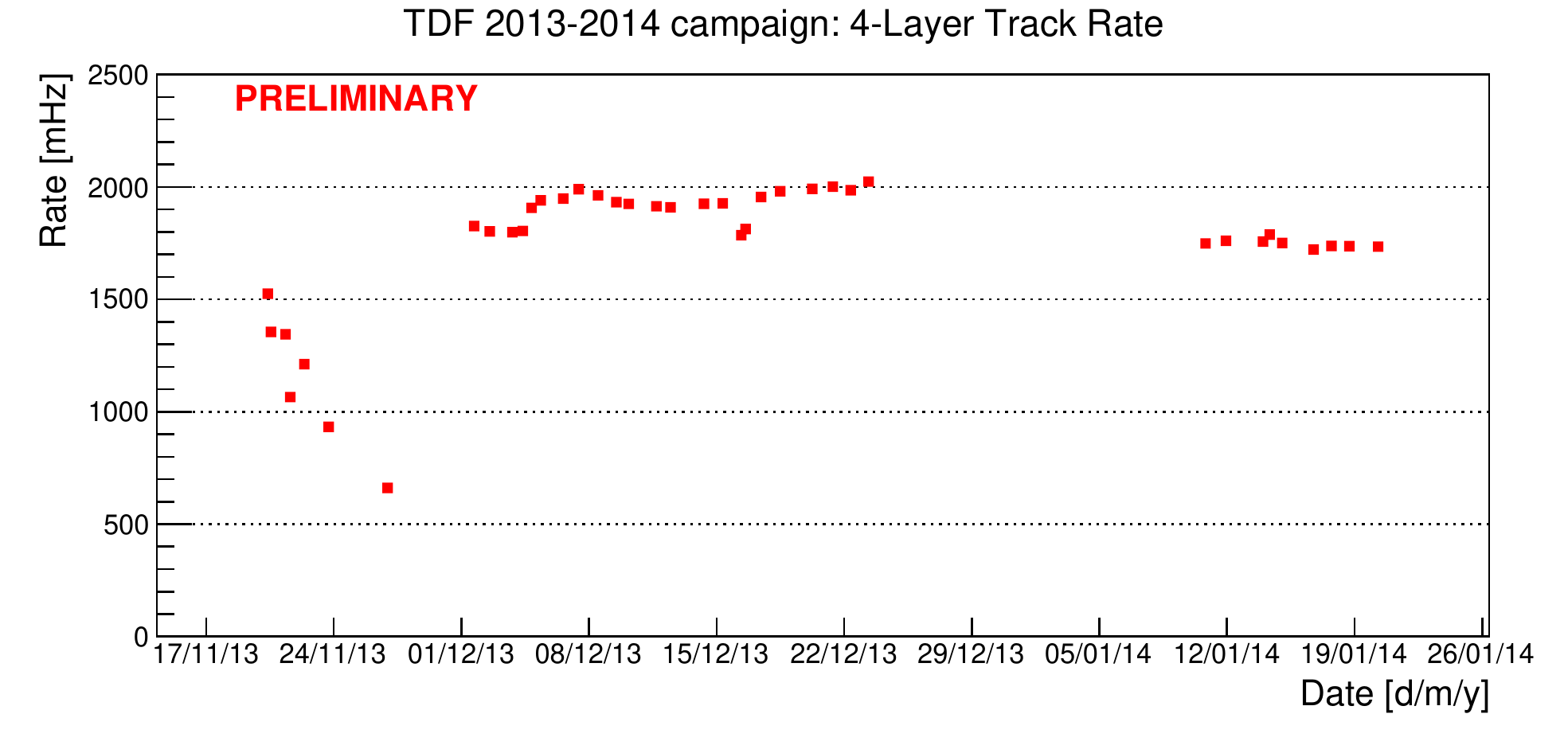}
  \caption{Efficiency (left) and rate (right) stabilities during the TDF campaign, from November 2013 to January 2014. On the efficiency plot, layer 1 is represented in black, layer 2 in red, layer 3 in green and layer 4 in blue.}
  \label{fig:effRateTDF2013}
\end{figure}

%\subsection{Efficiency}
Efficiency is one of the most important parameters in our case as a bias in this quantity will directly influence the muon flux estimation. It is calculated from the ratio of 3-layers tracks with an additional signal in the 4th layer matched to the track over the 3-layers tracks. Prior to the deployment, the chambers are tested in the laboratory. Since the thresholds are set globally for the 64 ASIC channels, the thresholds are scanned by making the efficiency versus high voltage and noise versus high voltage curves for each ASIC. Then a working HV point is defined for each chamber, as well as threshold values for each ASIC such as to maximise the efficiency while keeping the noise at an acceptable level. We aim to get a stable detector, which was the case during the TDF campaign from November 2013 to January 2014, as represented on figure~\ref{fig:effRateTDF2013} (left). Chambers from the first layer (in black) and from the last layer (in blue) look less efficient. This is mainly due to a geometrical effect as they are the two outside layers and give more constrain on the reconstructed direction of the track.\\

%\subsection{Rates stability}
The rate stability is represented on figure~\ref{fig:effRateTDF2013} (right) for the campaign at TDF from November 2013 to January 2014. It shows a stablility better than 6~\% after the end of the commissionning phase. The first points correspond to a HV scan.\\

\section{Data analysis}
\begin{figure}[htbp]
  \centering
  \includegraphics[width=0.45\textwidth]{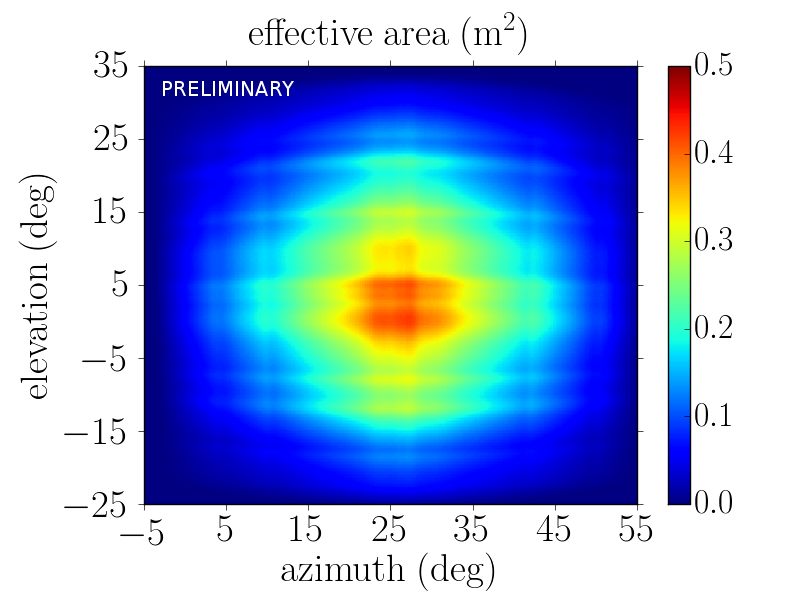}
  \includegraphics[width=0.45\textwidth]{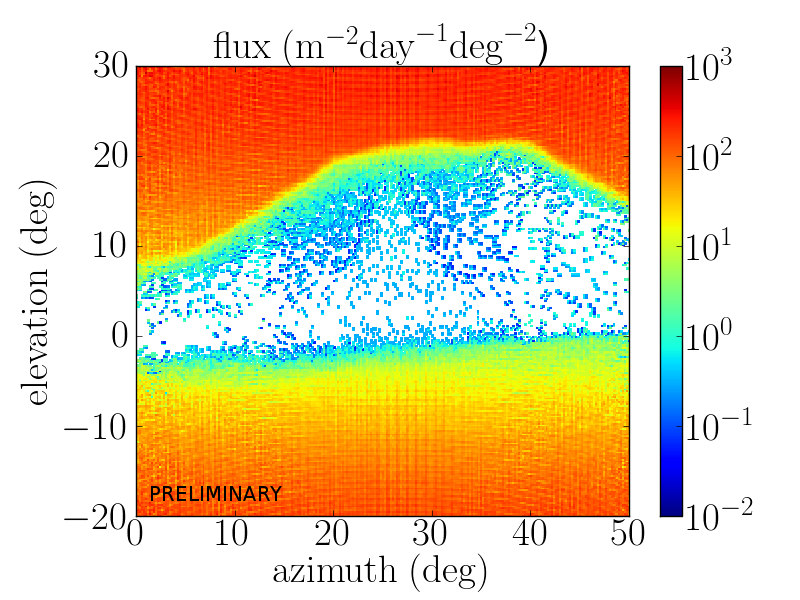}
  \caption{Effective surface (left) and reconstructed flux (right) in function of azimuth and elevation for the campaign in Col de Ceyssat from October 2015 to January 2016.}
  \label{fig:effSurfAflux}
\end{figure}

Using the performances of the detector its effective surface, defined as $S_{det}\times\varepsilon_{geom}\varepsilon_{illum}\varepsilon_{det}\varepsilon_{rec}\varepsilon_{sel}$, is calculated. $\varepsilon_{geom}$ is the geometrical acceptance, $\varepsilon_{illum}$ the illuminance correction factor, $\varepsilon_{det}$ the detector efficiency, $\varepsilon_{rec}$ the reconstruction efficiency and $\varepsilon_{sel}$ the selection efficiency. It is represented in function of the azimuth and the elevation on figure~\ref{fig:effSurfAflux} (left) for the campaign in Col de Ceyssat from October 2015 to February 2016. Our maximum effective surface for that campaign is about 0.4~m$^2$ for a 1~m$^2$ detector.

The incoming flux of atmospheric muons is then deduced and also represented on figure \ref{fig:effSurfAflux} (right). The shadow of the volcano can easily be identified on the positive elevations. The free sky in the opposite direction of the volcano is visible at negative elevations but also in the direction of the volcano, e.g. at elevations larger than 20$^\circ$. Statistics is largely lower inside the volcano.

\begin{figure}[htbp]
  \centering
  \includegraphics[width=0.35\textwidth]{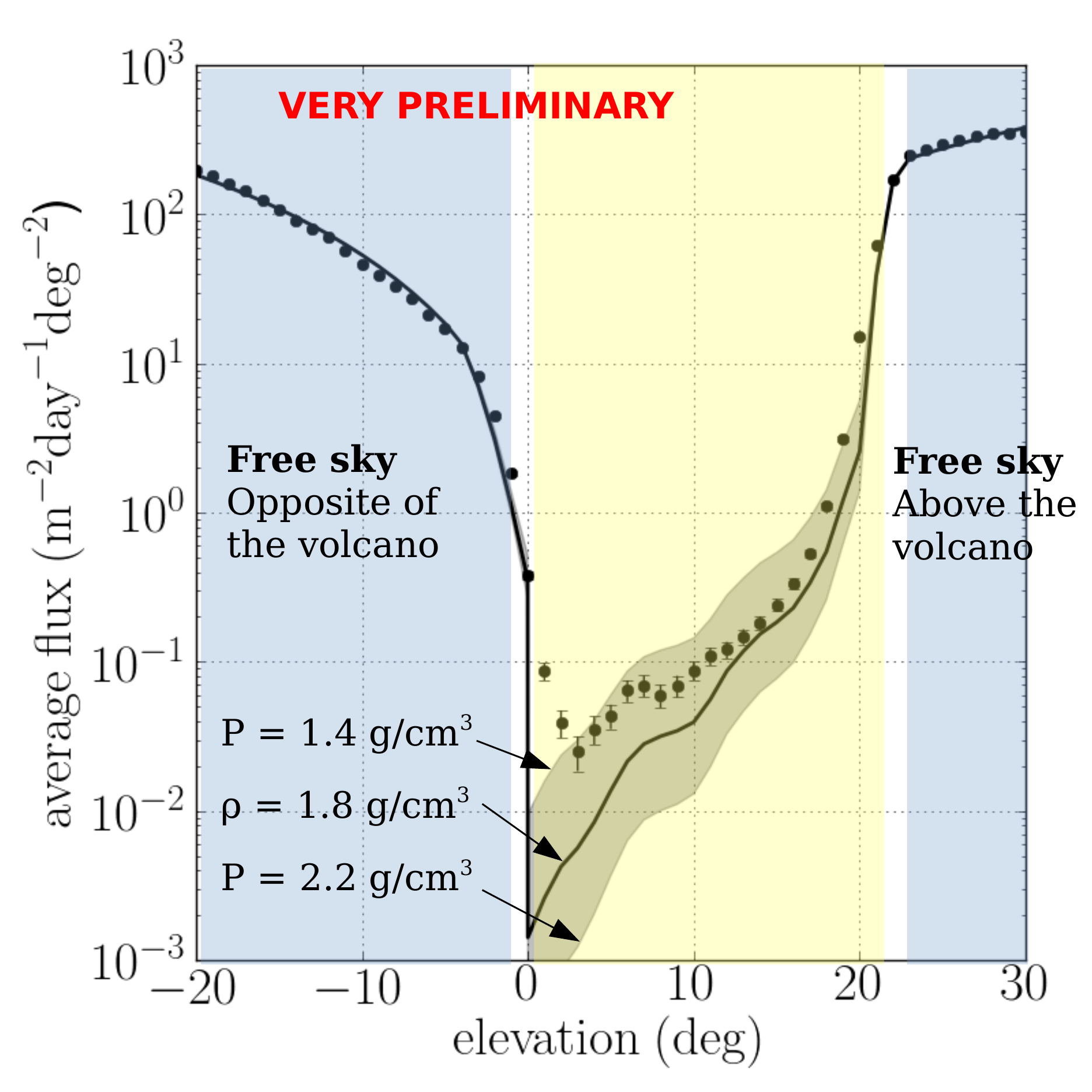}
  \includegraphics[width=0.45\textwidth]{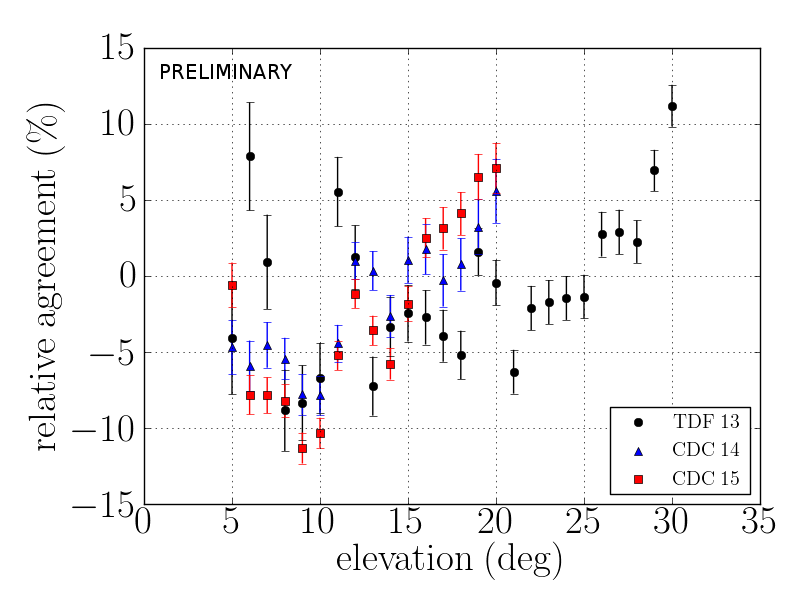}
  \caption{Reconstructed flux (dots) compared to prediction (left): the grey band corresponds to densities from 1.4~g.cm$^{-3}$ to 2.2~g.cm$^{-3}$; relative agreement between reconstucted flux and prediction for the free sky flux (right).}
  \label{fig:fluxComp}
\end{figure}

In order to reduce the statistical fluctuations, the incoming flux of atmospheric muons is represented in figure~\ref{fig:fluxComp} (left) integrated over an azimuth window of 20$^\circ$, centered on the volcano. The flux is then compared to our prediction, coming from a homemade Monte-Carlo code using the energy loss data provided by the Particle Data Group\cite{pdg2014,pdglive} and the parametrisation of the atmospheric flux spectrum provided in \cite{Chirkin2004}. This Monte-Carlo code was extensively tested against reference simulation codes like \textsc{Geant4}\cite{agostinelli2003} or MUM\cite{sokalski2001}. The tests included comparisons in simplified simulation configuration (e.g. survival probabilities after propagation through different materials of muons between MeV and PeV), but also comparisons with a detailed simulation based on \textsc{Geant4} and tailored for our experimental needs. As an example, the relative difference, the relative difference on the rock density predicted with our homemade Monte-Carlo or \textsc{Geant4} 10.01 is less than 0.3~\% after propagating through rock depths up to 10~km. At the same time, the speed gain compared with \textsc{Geant4} when propagating muons through 10~km of rock is $\times$~5000.

% the parametrisation of the atmospheric flux spectrum provided in \cite{Chirkin2004}. This Monte-Carlo code was extensively tested against general purpose simulation codes like \textsc{Geant4}\cite{agostinelli2003}, as well as codes dedicated to muon simulation like MUM\cite{sokalski2001}. The tests included comparisons in simplified simulation configuration (e.g. survival probabilities after propagation through different materials of muons between MeV and PeV), but also comparisons with a detailed simulation based on \textsc{Geant4} and tailored for our experimental needs. Whenever possible, the homemade simulation was also tested against data mainly coming from Particle Data Group\cite{pdg2014,pdglive}.
% As an example, the relative difference on the integrated muon rate ($\Delta\phi/\phi$) predicted by our  homemade Monte-Carlo and \textsc{Geant4} 10.01 is less than 4~\% after propagating through rock depths up to 10~km. At the same time, the gain speed compared with \textsc{Geant4} when propagating muons through 10~km of rock is $\times$~5000.

On figure~\ref{fig:fluxComp} (left), the density considered for the volcano in our prediction is 1.8~g.cm$^{-3}$, in agreement with the mean density derived from gravimetric data and measured rock samples. The grey area represents the density range of these surveys: 1.4 to 2.2~g.cm$^{-3}$. The relative agreement between prediction and the measured flux in the free sky regions (in the blue areas), shown on figure~\ref{fig:fluxComp} (right), is within 10~\% for the moment. It will be improved in next months, mainly by better understanding and accounting for the detector performance.

\section{Developments and conclusion}
Various improvements are ongoing on our detector. Our goal is to improve the spatial and the temporal resolution by a factor ten by using multi-gap GRPCs. In order to get a more portable detector, easier to use in-situ, we are also working on the reduction of the cost, the electrical consumption and the complexity, using strips for example. A new scheme for gas circulation with new inlets and outlets is also under test, mainly to reduce gas consumption.

The campaigns on Puy de D\^ome showed that GRPCs can be successfully used for muography. GRPCs offer a detector with good time and space resolution for a reasonable price. New detectors, as by example, scintillators using micro-fibers, offer the required spatial resolutions, but the price per unit of instrumented surface is still prohibitive. We are pursuing our studies in order to better understand our telescope and give a quantitative result on the Puy de D\^ome volcano. Some of the major points are the fine understanding of the detector and reconstruction efficiencies. Data analysis is currently ongoing and more quantitative results are expected soon.

%Such resolution can now be achieve with scintillators using micro-fibers, but this is new and more expensive for a big detection area.

%GRPCs showed that they can be successfully used for muography and more quantitative results are expected soon.
%% \section{Appendix}
%% Please always give a title also for appendices.

\acknowledgments
The TOMUVOL collaboration acknowledges funding from the University Blaise Pascal of Clermont-Ferrand, CNRS, R\'egion Auvergne and Conseil G\'en\'eral du Puy-de-D\^ome and the AIDA 2020 program (grant agreement number 654168). During the data taking campaigns, the TOMUVOL detector was kindly hosted in a building belonging to TDF Rh\^one Auvergne in 2013, thanks to Luc Lecoeuvre, head of the TDF Housing Wealth department. The author of this article was supported by Clervolc Labex program (ANR-10-LAB-0006). This is Clervolc contribution 204.

%% \paragraph{Note added.} This is also a good position for notes added
%% after the paper has been written.

This is an author-created, un-copyedited version of an article published in JINST {\bf 11} C06009. IOP Publishing Ltd is not responsible for any errors or omissions in this version of the manuscript or any version derived from it. The Version of Record is available online at 10.1088/1748-0221/11/06/C06009.



\begin{thebibliography}{99}

\bibitem{brown2015}
S.K. Brown et al., \emph{Global volcanic hazards and risk: Technical background paper for the UN-ISDR Global Assessment Report on Disaster Risk Reduction 2015}, Global Volcano Model and the International Association of Volcanology and Chemistry of the Earth's Interior (2015).

\bibitem{revil2008}
A. Revil et al., \emph{Inner structure of La Fossa di Vulcano (Vulcano Island, southern Tyrrhenian Sea, Italy) revealed by high-resolution electric resistivity tomography coupled with self-potential, temperature, and CO$_2$ diffuse degassing measurements.}, \emph{J. Geophys. Res.} {\bf 113} (2008) B07207.

\bibitem{li1998}
Y. Li and D. Oldenburg, \emph{3-D inversion of gravity data}, \emph{Geophysics} {\bf 63} (1998) pg. 109 - 119

\bibitem{gailler2009}
L. Gailler et al., \emph{Gravity structure of Piton de la Fournaise volcano and inferred mass transfer during the 2007 crisis}, \emph{J. Volcanol. Geotherm. Res.} {\bf 184} (2009) pg. 31 - 48.

\bibitem{gailler2012}
L. Gailler and J.-F. L\'enat, \emph{Architecture of La R\'eunion inferred from geophysical data}, \emph{Bull. Volcanol.} {\bf 221-222} (2012) pg. 83 - 98.

\bibitem{lees2007}
J. M. Lees, \emph{Seismic tomography of magmatic systems}, \emph{J. Volcanol. Geotherm. Res.} {\bf 167(1-4)} (2007) pg. 37 - 56.

\bibitem{alvarez1970}
L. W. Alvarez et al., \emph{Search for hidden chambers in the pyramids}, \emph{Sience} {\bf 167(3919)} (1970) pg. 832 - 839.

\bibitem{nagamine1995}
K. Nagamine et al., \emph{Method of probing inner-structure of geophysical substance with the horizontal cosmic-ray muons and possible application to volcanic eruption prediction}, \emph{Nucl. Instrum. Methods Phys. Res., Sect. A} {\bf 336} (1995) pg. 585 - 595.

\bibitem{tanaka2007}
H. K. M. Tanaka et al., \emph{Development of an emulsion imaging system for cosmic-ray muon radiography to explore the internal structure of a volcano, Mt. Asama}, \emph{Nucl. Instrum. Methods Phys. Res., Sect. A} {\bf 575} pg. 489 - 497.

\bibitem{marteau2012}
J. Marteau et al., \emph{Muon tomography applied to geosciences and volcanology}, \emph{Nucl. Instrum. Methods Phys. Res., Sect. A} {\bf 695} pg. 23 - 28.

\bibitem{fehr2012}
F. Fehr, \emph{Density imaging of volcanos with atmospheric muons}, \emph{J. Phys. Conf. Ser.} {\bf 375} (2012) 052019.

\bibitem{tanaka2014}
H. K. M. Tanaka et al., \emph{Radiographic visualization of magma dynamics in an erupting volcano}, \emph{Nat. Commun.} {\bf 5} (2014) 3381.

\bibitem{beaulieu2015}
G. Beaulieu et al., \emph{Construction and commissioning of a technological prototype of a high-granularity semi-digital hadronic calorimeter}, \emph{JINST} {\bf 10} (2015) 10 P10039.

\bibitem{callier2014}
S. Callier et al., \emph{ROC chips for imaging calorimetry at the International Linear Collider}, \emph{JINST} {\bf 9} (2014) 02 C02022.

\bibitem{gonzalez2005}
D. Gonz\'{a}lez-D\'{i}az et al., \emph{The effect of temperature on the rate capability of glass timing RPCs}, \emph{Nucl. Instrum. Methods Phys. Res., Sect. A} {\bf 555} (2005) pg. 72 - 79.

\bibitem{pdg2014}
K. A. Olive et al., \emph{Review of Particle Physics}, \emph{Chin. Phys. C} {\bf 38} (2014) 090001.

\bibitem{pdglive}
PDG Live, \emph{Atomic and Nuclear Properties of Materials}, \url{http://pdg.lbl.gov/2015/AtomicNuclearProperties/index.html}

\bibitem{Chirkin2004}
D. Chirkin, \emph{Fluxes of Atmospheric Leptons at 600 GeV - 60 TeV}, arxiv:hep-ph/0407078 (2004)

\bibitem{agostinelli2003}
S. Agostinelli et al., \emph{Geant4-a simulation toolkit}, \emph{Nucl. Instrum. Methods Phys. Res., Sect. A} {\bf 506} (2003) pg. 250 - 303.

\bibitem{sokalski2001}
I. A. Sokalski et al., \emph{MUM: Flexible precise Monte Carlo algorithm for muon propagation through thick layers of matter}, \emph{Phys. Rev. D} {\bf 64} (2001) 074015.


%% \bibitem{b}
%% Author, \emph{Title},, \emph{J. Abbrev.} {\bf vol} (year) pg.
%% arxiv:1234.5678.

%% \bibitem{c}
%% Author, \emph{Title},
%% Publisher (year).


% Please avoid comments such as "For a review'', "For some examples",
% "and references therein" or move them in the text. In general,
% please leave only references in the bibliography and move all
% accessory text in footnotes.

% Also, please have only one work for each \bibitem.


\end{thebibliography}
\end{document}